\documentclass[11pt]{article}
\usepackage{moriond,epsfig}

\def\be{\begin{equation}}
\def\ee{\end{equation}}
\def\bea{\begin{eqnarray}}
\def\eea{\end{eqnarray}}

\def\msun{{\rm ~M}_{\odot}}

\def\mpy{{\rm ~M}_{\odot} {\rm ~yr}^{-1}}
\def\zsun{{\rm ~Z}_{\odot}}

\begin{document}
\vspace*{4cm}
\title{THE COALESCENCE RATES OF DOUBLE BLACK HOLES}

\author{BELCZYNSKI K.$^{1,2}$, BULIK, T.$^{1}$, DOMINIK., M.$^{1}$, PRESTWICH,
A.$^{3}$}

\address{
$^{1}$ Astronomical Observatory, University of Warsaw, Al.
            Ujazdowskie 4, 00-478 Warsaw, Poland\\
$^{2}$ Center for Gravitational Wave Astronomy, University of Texas at
            Brownsville, Brownsville, TX 78520, USA\\
$^{3}$ Harvard-Smithsonian Center for Astrophysics, Cambridge, MA 02138, USA
}

\maketitle\abstracts{
We present the summary of the recent investigations of double black hole
binaries in context of their formation and merger rates. In particular we
discuss the spectrum of black hole masses, the formation scenarios in the
local Universe and the estimates of detection rates for gravitational
radiation detectors like LIGO and VIRGO. Our study is based on observed
properties of known Galactic and extra-galactic stellar mass black holes and 
evolutionary predictions. We argue that the binary black
holes are the most promising source of gravitational radiation. 
}

\section{Population Synthesis Coalescence Rates}

We employ the {\tt StarTrack} population synthesis code (Belczynski et al.
2002, 2008) to perform several Monte Carlo simulations of binary evolution 
with a range of metallicity. We base the calculations on recent results
from the Sloan Digital Sky Survey observations (Panter et al. 2008)
indicating ($\sim 300,000$ galaxies) that recent star
formation (within the last 1 billion years) is bimodal: half the stars form
from gas with high amounts of metals (solar metallicity), and the other half
form with small contribution of elements heavier than Helium ($\sim 10-30\%$
solar). Additionally, we use the recent estimates of mass loss rates 
producing much heavier stellar black holes than previously expected ($\sim
30-80 \msun$; Belczynski et al. 2010a). 
The results of these calculations were presented for the first time by 
Belczynski et al. (2010b).
We have evolved a population of 2 million massive binary stars, and
investigated the formation of close  
double compact objects: double neutron stars (NS-NS), double black hole
binaries (BH-BH), and mixed systems (BH-NS). Our modeling utilizes updated 
stellar and binary physics, including results from supernova simulations 
(Fryer \& Kalogera 2001) and compact object formation (Timmes et al. 1996), 
incorporating elaborate mechanisms for treating stellar interactions like mass 
transfer episodes and tidal synchronization and circularization. We put 
special emphasis on the common envelope 
evolution phase, which is crucial for close double compact 
object formation as the attendant mass transfer allows for efficient 
hardening of the binary. This orbital contraction can be sufficiently 
efficient to cause the individual stars in the binary to coalesce
and form a single highly rotating object, thereby aborting further binary
evolution and preventing the formation of a double compact object. Due to
significant radial expansion, stars crossing the Hertzsprung gap (HG) very
frequently initiate a common envelope phase. HG stars do not have a clear
entropy jump at the core-envelope transition (Ivanova \& Taam 2004); if such
a star overflows its Roche lobe and initiates a common envelope phase, the
inspiral is expected to lead to a coalescence (Taam \& Sandquist 2000). In
particular, it has been estimated that for a solar metallicity environment
(e.g., our Galaxy), properly accounting for the HG gap may lead to a reduction 
in the merger rates of BH-BH binaries by $\sim 2-3$ orders of magnitude 
(Belczynski et al. 2007). The details of the common envelope phase are not yet 
fully understood, and thus in what follows we consider two models, one which 
does not take into account the suppression (optimistic model: A), and one that 
assumes the maximum suppression (pessimistic model: B).

\begin{table}[t]
\caption{Galactic Merger Rates [Myr$^{-1}$]}
\vspace{0.4cm}
\begin{center}
\begin{tabular}{lccc}
\hline
\hline
& $\zsun$ & $0.1\ \zsun$ & $\zsun$ + $0.1\ \zsun$  \\                
 Type & ($100\%$)    & ($100\%$) & ($50\%$ + $50\%$) \\
\hline
NS-NS               &  40.8 (14.4)  &  41.3 (3.3)  &  41.1 (8.9) \\
BH-NS               &  3.2  (0.01)  &  12.1 (7.0)  &   7.7 (3.5) \\
BH-BH               &  1.5 (0.002)  &  84.2 (6.1)  &  42.9 (3.1) \\
TOTAL               &  45.5 (14.4)  &  138 (16.4)  &  91.7 (15.4) \\  
\hline
\end{tabular}
\end{center}
\end{table}

\begin{table}[t]
\caption{LIGO/VIRGO Detection Rates [yr$^{-1}$]}
\vspace{0.4cm}
\begin{center}
\begin{tabular}{llccc}
\hline
\hline
Sensitivity & & $\zsun$ & $0.1\ \zsun$ & $\zsun$ + $0.1\ \zsun$  \\
($d_{\rm 0,nsns}$=) & Type & ($100\%$)    & ($100\%$) & ($50\%$ + $50\%$)\\

        &NS-NS                                     &  0.01 (0.003)  &  0.01 (0.001)    &  0.01 (0.002)  \\
18 Mpc  &BH-NS                                     &  0.007 (0.00002)  &  0.04 (0.02)  &  0.02 (0.01) \\
        &BH-BH                                     &  0.02 (0.00005)   &  9.9 (0.1)    &   4.9 (0.05) \\
        &TOTAL                                     &  0.03 (0.003)      & 10.0 (0.1)   &  5.0 (0.06) \\
        &&&&\\
\hline
\end{tabular}
\end{center}
\end{table}

The results are presented in Table 1 (Galactic merger rates) and 2
(LIGO/VIRGO detection rates). 
In Table 1 the rates are calculated for a Milky Way type galaxy (10 Gyr of continuous star
formation at a rate of $3.5 \mpy$), with the assumption that all
stars have either solar metallicity or $10\%$ solar, or a 50-50 mixture of both
types of stars. The rates are presented for the optimistic model (A) where
progenitor binaries survive through the common envelope phase, while the results
in parentheses represent the pessimistic model (B), where the binaries do not
survive if the phase is initiated by a Hertzsprung gap star.  
In table 2 the detection rates are given for model A (B) for a given 
sensitivity of LIGO/VIRGO instrument. 
Sensitivity is defined as the sky and angle averaged distance horizon for
detection of a NS-NS inspiral. 
The rates are given for a local Universe consisting of only solar composition
stars (unrealistically high), $0.1 \zsun$ stars (unrealistically low) and for
a 50-50 mixture of the above (realistic local Universe;~Panter et al. 2008). The sensitivity 
of $d_{\rm 0,nsns}=18\,\mbox{Mpc}$ corresponds to the
expected initial LIGO/VIRGO detector. 

The results show two clear trends. First, the rates are generally larger for
model A than B. This is the direct consequence of our assumptions on common
envelope outcome in both models as mentioned earlier and discussed in detail
by Belczynski et al. (2007). Since black hole progenitors are the most
massive stars and thus experience the most dramatic expansion (CE mergers in
model B) the BH-BH rates are affected in the largest extent. 
Second, we note that the rates are higher for the low metallicity model
($Z=0.1 \zsun$) as compared with high metallicity model ($Z=\zsun$). 
The major reason behind this trend is the smaller radii of stars at low
metallicity. This directly leads back to CE evolution; the smaller the
radius of a given star the later in evolution the star overflows its Roche
lobe. Thus for low metallicity, massive stars tend to initiate CE phase after
HG, and so they have a chance of surviving this phase and
forming a double compact object independent of assumed model of CE evolution.  
The increase of rates with decreasing metallicity is additionally connected 
with the fact that low metallicity stars experience low wind mass loss and
thus form more massive compact objects. This leads to a shorter merger times
and higher merger rates. 

Had the initial configuration of LIGO/VIRGO instruments reached its design
sensitivity of $d_{\rm 0,nsns} 18$ Mpc for its entire lifetime
(averaged horizon for NS-NS merger) we would be able
to exclude model A from our considerations. This model generates about 5
BH-BH inspirals per year within this horizon (of course the actual horizon
for BH-BH detection was accordingly extended with the increased mass of each
BH-BH merger). So far there was no report of detection in LIGO/VIRGO
data so one would be tempted to exclude this model from further considerations.
However, the time averged sensitivity of the last LIGO/VIRGO run (S5, the
most recently relesed) has reached only about $d_{\rm 0,nsns} \approx
9$ Mpc. Therefore, the rates should be decreased by factor $(18/9)^3=8$ and 
the expected detection rate for BH-BH binaries would drop below $1$
yr$^{-1}$ and consequently model A cannot be yet excluded.

\section{Empirical Coalescence Rates}

The optical followup of X-ray sources revealed
the nature of several X-ray binaries in the galaxies in the Local Group.
Two objects: IC10 X-1 and NGC300 X-1 are of particular interest.
The identification of optical counterparts and their spectroscopy 
allowed to estimate the properties of these two binaries.
Both host massive black holes on a tight orbit with
WR stars. Both reside in low metallicity environments
(Crowther et al. 2007, Crowther et al. 2010, Prestwich et al. 2007,
Silverman and Filipenko 2008).
In the future the accretion in these binaries will continue and the WR stars will
loose mass through stellar winds. The typical lifetime of 
the WR stars in such systems is from 100 to 300~kyrs.
After that time the WR stars will explode as supernovae 
leading to formation of a BH, or a NS
in the case of extremely large mass loss. 
The systems will most likely survive the explosions and remain 
bound since the current orbital velocities are above $500$km\,s$^{-1}$.
Both systems will end up as binary black holes 
in a few hundred thousand years.

Given the  estimate of the future evolution of the 
two binaries: IC10 X-1 and NGC300 X-1, we estimate the formation 
rate of such binaries.The estimated merger time 
is smaller than the Hubble time. Therefore 
assuming that the star formation rate was constant
the merger rate of the 
binary black holes formed from such systems will be the same as their 
formation rate. 
For each system we estimate the volume in which it is detectable. 
This is determined by the possibility of measuring the radial velocity curve,
which can be done up to the distance of $r\approx 2$Mpc, thus $V_{obs}=4\pi
r^3/3$. Each binary 
was detected only because of its X-ray radiation, thus the observability 
is proportional to the X-ray active phase. The formation rate of each binary 
can be approximated as: $R=(V_{obs} t_{obs})^{-1}$.
A detailed statistical analysis is presented in 
Bulik, Belczynski and Prestwich (2010).
We present the  probability distributions of the formation
and merger rates of the binary black holes corresponding to each 
binary IC10 X-1 and NGC300 X-1 in Figure~1.
The thick line in Figure~1 represents the probability density 
of the sum of the two rates. This 
calculation implies a merger rate density of
 ${\cal R} =  0.36^{+0.50}_{-0.26}$Mpc$^{-3}$Myr$^{-1}$.
For the time averaged the sensitivity range of LIGO and VIRGO 
to binary black holes coalescence of $\approx 100$Mpc, this implies
the expected detection rate around one per year. This 
is in a striking agreement with the population synthesis results.

\begin{figure}
\begin{center}
\includegraphics[width=0.5\textwidth]{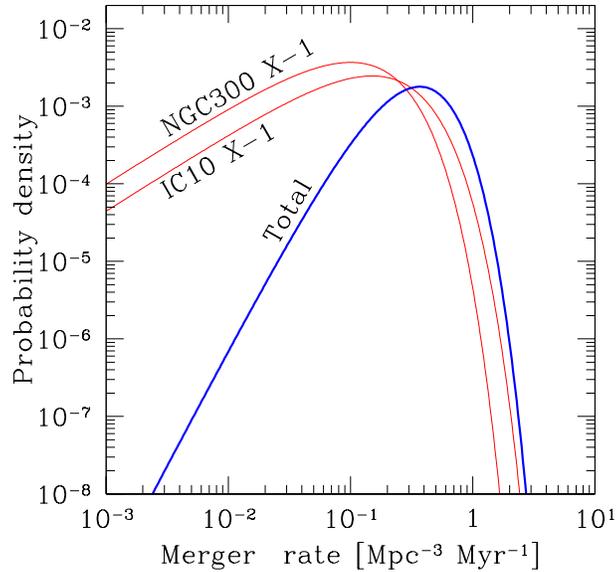}
\end{center}
\caption{The probability density of the binary black hole 
merger rate density. We present separately the contributions of IC10 X-1 and
NGC300 X-1 and the total rate.}
\end{figure}

\section{Conclusions}

Both theoretical simulations and empirical estimate
indicate that detection rates of BH-BH binaries are significantly higher
than other double compact objects (NS-NS and BH-NS). 
The population synthesis predictions for our realistic model of local
Universe with a mixture of high and low metallicity stars results in about
100 BH-BH detections per 1 NS-NS detection by LIGO/VIRGO. 
The empirical estimate presented here for BH-BH detection
rate based on the observed extra-galactic BH binaries is $\approx 1$
yr$^{-1}$, again much higher than the corresponding empirical detection rate
for NS-NS inspiral ($\approx 0.06$ yr$^{-1}$, Kim, Kalogera \& Lorimer 2006). 
Thus it is likely that the existing LIGO/VIRGO data contains a coalescence 
signal that may be discovered with a more elaborate reanalysis.

\section*{Acknowledgments}
Authors acknowledge support from MSHE grants N N203 302835 and N N203 404939.


\section*{References}
\begin{thebibliography}{99}

\bibitem{2004ApJ...601L..67B} Bauer, F.~E., \& Brandt, W.~N.\ 2004,
ApJ, 601, L67
\bibitem{Bel02} Belczynski, K., Kalogera, V., \& Bulik, T.\ 2002, ApJ, 572, 407
\bibitem{Bel07} Belczynski, K., et al.\ 2007, ApJ, 662, 504
\bibitem{Bel08} Belczynski, K., et al.\ 2008, ApJ Sup., 174, 223
\bibitem{Bel10a} Belczynski K., et al.\ 2010a, ApJ, 714, 1217
\bibitem{Bel10b} Belczynski, K., et al.\ 2010b, ApJ, 715, L138
\bibitem{bbp} Bulik, T., Belczynski, K., Pretwich, A., 2011, ApJ, 730, 140
\bibitem{2010MNRAS.403L..41C} Crowther, P.~A., et al.\ 2010, MNRAS, 403, L41
\bibitem{2007A&A...469L..31C} Crowther, P.~A., Carpano, S., Hadfield,
         L.~J., \& Pollock, A.~M.~T.\ 2007, A\&A, 469, L31
\bibitem{Fry} Fryer, C., \& Kalogera, V.\ 2001, ApJ, 554, 548
\bibitem{Iva} Ivanova, N., \& Taam, R. E.\ 2004, ApJ, 601, 1058
\bibitem{Kim} Kim, C., Kalogera, V., \& Lorimer, D.\ 2006, New Astronomy Rev. 54, 
148 
\bibitem{Pan} Panter B., et al.\ 2008, MNRAS, 391, 1117
\bibitem{2007ApJ...669L..21P} Prestwich, A.~H., et
         al.\ 2007, ApJ, 669, L21
\bibitem{2008ApJ...678L..17S} Silverman, J.~M., \& Filippenko, A.~V.\
2008, ApJ, 678, L17
\bibitem{Tam} Taam, R. E., \& Sandquist, E. L. 2000, ARA\&A, 38, 113
\bibitem{Tim} Timmes, F., Woosley, S., \& Weaver, T.\ 1996, ApJ, 457, 834


\end{thebibliography}
\end{document}